\documentclass[aps,prb,amsmath,amssymb,floatfix,twocolumn,amsmath,superscriptaddress,nofootinbib,letterpaper]{revtex4-1}
\usepackage{multirow}
\usepackage{bbold}
\usepackage{color}
\usepackage{mathrsfs}
\usepackage{hyperref}
\usepackage[normalem]{ulem}
\usepackage{bm}

\usepackage{amsfonts, relsize, color}
\usepackage{graphics}
\usepackage{graphicx}
\usepackage{hyperref}
\usepackage{color}
\usepackage{siunitx}
\usepackage{subcaption}
\begin{document}
\title{\bf Discontinuous evolution of the structure of stretching polycrystalline graphene}

\author{Federico D'Ambrosio*}
\affiliation{Department of Information and Computing Sciences,
Utrecht University, Princetonplein 5,
3584 CC Utrecht, The Netherlands}

\author{Vladimir Juri\v ci\' c}
\affiliation{Nordita, KTH Royal Institute of Technology and Stockholm University, Roslagstullsbacken 23,  10691 Stockholm,  Sweden}

\author{Gerard T. Barkema}
\affiliation{Department of Information and Computing Sciences,
Utrecht University, Princetonplein 5,
3584 CC Utrecht, The Netherlands}

\begin{abstract}
Polycrystalline graphene has an inherent tendency to buckle, i.e. develop out-of-plane, three-dimensional structure. A force applied to stretch a piece of polycrystalline graphene influences the out-of-plane structure. Even if the graphene is well-relaxed, this happens in
non-linear fashion: occasionally, a tiny increase in stretching force induces a significant displacement, in close analogy to avalanches, which in turn can create vibrations in the surrounding medium. We establish this effect in computer simulations: by continuously changing the strain, we follow the displacements of the carbon atoms that turn out to exhibit a discontinuous evolution. Furthermore, the displacements exhibit a hysteretic behavior upon the change from low to high stress and back. These behaviors open up a new direction in studying dynamical elasticity of polycrystalline quasi-two-dimensional systems, and in particular the implications on their mechanical and thermal properties.
\end{abstract}

\maketitle

\emph{Introduction.} Graphene, a crystal of carbon atoms arranged in a honeycomb lattice,
shows a plethora of exotic mechanical and electronic properties, and emerged as a paradigmatic
example of a crystalline membrane embedded in the three-dimensional space \cite{CastroNeto2009,Geim2009,Nair2012,Smith2013,Dolleman2016,Cartamil-Bueno2016,Lee2008,Milowska2013}.
In particular, it exhibits an intrinsic  tendency to spontaneously buckle when it is polycrystalline, i.e, when it features many crystalline domains, or due to the presence of lattice defects, such as disclinations and Stone-Wales defects \cite{Fasolino2007,banhart2010structural,zhang2014defects}.
This effect arises due to the competition of stress, introduced either by mismatch between the domains or by the defects, and the tendency of the membrane to relieve the stress by bending, which yields a rather rich landscape of configurations for a relaxed membrane \cite{Witten2007}. In addition, a graphene sheet can experience external stress, for instance when graphene is held by clamps which exert a pulling force, which further enriches the landscape of the ground state configurations.

It is well known how a polycrystalline or defected graphene sheet relaxes when subjected to a constant, static external stress~\cite{Si2016}, but the case of dynamic strain remained rather unexplored. The latter case can be experimentally relevant as external perturbations creating strain are in reality time-dependent. In particular, when external stress is gradually applied to a polycrystalline graphene sample, it is of a fundamental and practical importance to establish whether  the change  of the shape of the graphene membrane is continuous or it follows a discontinuous path in this rather complex configuration space.

In this paper, by performing computer simulations, we show that the evolution of the shape of a polycrystalline graphene membrane is discontinuous: as the stress is uniformly increasing, occasionally a tiny increase in stretching force induces a significant displacement, analogous to avalanches, see Fig.~1 and Supplemental Material at [URL will be inserted by publisher] for a video of the evolution of the sample due to the external stress that highlights this avalanche-like behavior, which, as a result, can create vibrations in the surrounding medium. Furthermore, if the stretching force is then decreased again to its starting value, the system does not follow the same path in the configuration space, but rather exhibits a hysteretic behavior while undergoing a cycle in the configuration space, see Fig.~2. The change of the profile of the sample takes place through the creation and annihilation of ridges and vertices, with the elastic energy concentrated in these defects, as shown in Fig.~3. Our findings should have implications for the dynamical elasticity in polycrystalline and defected sheets of graphene, and, in particular, we expect that various elastic moduli, phonon density of states, and thermal conductivity will be affected.

\emph{Model.}
We opted for the empirical graphene potential introduced in Ref.~\cite{Jain2015}, which is based on Kirkwood's potential \cite{Kirkwood1939}. This potential has been used, for instance, for studying the long-range relaxation of structural defects \cite{Jain2015}, for probing crystallinity of graphene samples via their vibrational spectrum \cite{Jain2015a}, for the study of twisted and buckled bilayer graphene \cite{Jain2017} as well as of the shape of a graphene nanobubble \cite{Jain2017a}. In detail, for a two-dimensional hexagonal network for which out-of-plane deformations are also allowed, this potential can be written as
\begin{eqnarray}\label{eq:elastic-energy}
E_0 &=& \frac{3}{16}\frac{\alpha}{d^2} \sum_{i,j} \left(r_{ij}^2 - d^2\right)^2\nonumber\\
&+&\frac{3}{8} \beta d^2 \sum_{j,i,k} \left( \theta_{j,i,k} - \frac{2 \pi}{3}\right)^2 + \gamma \sum_{i,jkl} r_{i,jkl}^2
\end{eqnarray}
with $d=\SI{1.420}{\angstrom}$, the ideal bond length for graphene. A two-body bond stretching energy contribution is parametrized by 
 $\alpha = \SI{26.060}{\electronvolt \per \angstrom^2}$, $\beta=\SI{5.511}{\electronvolt \per \angstrom^2}$ controls the bond shearing contribution, while $\gamma=\SI{0.517} {\electronvolt \per \angstrom^2}$ corresponds to the energy cost for the out-of-plane deformation of the graphene membrane \cite{Jain2015}. 
Since the periodic boundary conditions apply, we can represent the contribution to the elastic energy due to external stress by an additional term of the form
\begin{equation}
\label{eq:ESigma}
E_\sigma = E_0 - \sigma L_x L_y = E_0 - \sigma S
\end{equation}
where $L_x$ and $L_y$ are the lateral dimensions of the two-dimensional periodic box, $S$  its surface and $\sigma$ the parameter that controls the strength of the stretching force. 

We first generate a polycrystalline flat sample by following the procedure described in Ref.~\cite{Jain2015}. A sample is allowed to relax to a lower energy configuration following the Fast Inertial Relaxation Engine (FIRE) algorithm \cite{Bitzek2006}. The values of the parameters in this algorithm ($N_{min}$, $f_{inc}$, $f_{dec}$, $\alpha_{start}$ and $f_{\alpha}$) are taken as suggested in Ref.~\cite{Bitzek2006}. Further relaxation of the sample requires topological changes in the network, which we perform through so called bond transpositions \cite{Wooten1985}. In this procedure, four connected atoms are selected, two bonds are then broken and reassigned between them. In the implementation we use an improved algorithm \cite{Barkema2000} that avoids complete relaxation before rejecting a bond transposition. After a random bond switch the complete sample is relaxed and the new configuration is accepted with the Metropolis probability 

\begin{equation}
P = \min \left\{ 1,\, \exp \left( \frac{E_b - E_a}{k_B T}\right)\right\}
\end{equation}
which includes the effect of thermal fluctuations: a move that increases the energy of the sample might be accepted, with a higher probability as temperature increases. However, note that no bond transposition is allowed while we perform the cycle in the configuration space. Temperature does therefore partially influence the evolution of the sample, and as long as there are still topological lattice defects around which the stress can accumulate, the discontinuous evolution persists.

Once this flat sample is sufficiently relaxed, every atom is placed at a random non-zero distance from the two-dimensional plane and allowed to relax to a buckled three-dimensional configuration. As topological defects increase the elastic energy of the sample, both globally and locally, they cause stress in the material that, once the material is allowed to relax in a three-dimensional configuration, is released by buckling \cite{Jain2015}. Using such configuration of the graphene membrane, we perform a cycle in the configuration space by manipulating the stretching force  $\sigma$ in the empirical potential, given by Eq.~(\ref{eq:ESigma}), while the topology of the sample is kept fixed. During this process, we increase the stress in steps of size $\Delta \sigma$, and, after each step, the sample is allowed to relax. Once the value of the maximum stress $\sigma_{max}$ is reached, it is decreased in the same way. In this work, $\Delta \sigma = \SI{8e-9}{\atomicmassunit\per\micro\second\squared}$ and $\sigma_{max} = \SI{8e-6}{\atomicmassunit\per\micro\second\squared}$, with $\si{\atomicmassunit}$ the atomic mass unit. 

Between each stretching step, the atoms in the graphene membrane translate to a different position, and the relevant translations are only the non-affine (intrinsic) ones, which are defined in the reference frame fixed to the sample itself, and therefore they are not related to the expansion or contraction of the sample. In order to characterize the evolution of the graphene sample under the applied external stress, we thus use the non-affinity parameter defined in Ref.~\cite{Huisman2010} as 
\begin{equation}\label{eq:non-affine-parameter}
A =\frac{\langle\left({\bf r}_i - {\bf r}_{i,A}\right)^2\rangle}{L_x L_y}
\end{equation}
where ${\bf r}_i$ is the in-plane position of the atom $i$, while  ${\bf r}_{i,A}$ is its expected position due to the expansion of the reference frame and $\langle\cdot\rangle$ is the average on all the atoms. 

We can also extract the local energy distribution at a given moment with distinct energy contributions for the different terms in the total elastic energy, given by Eq.~(\ref{eq:elastic-energy}). This is computed by dividing the two body energy of every bond equally between the two atoms and assigning the three body and out of plane energy to the central atom.

\begin{figure}[t]
    \centering
    \begin{subfigure}[t]{0.45\textwidth}
        \centering
        \includegraphics[width=\textwidth]{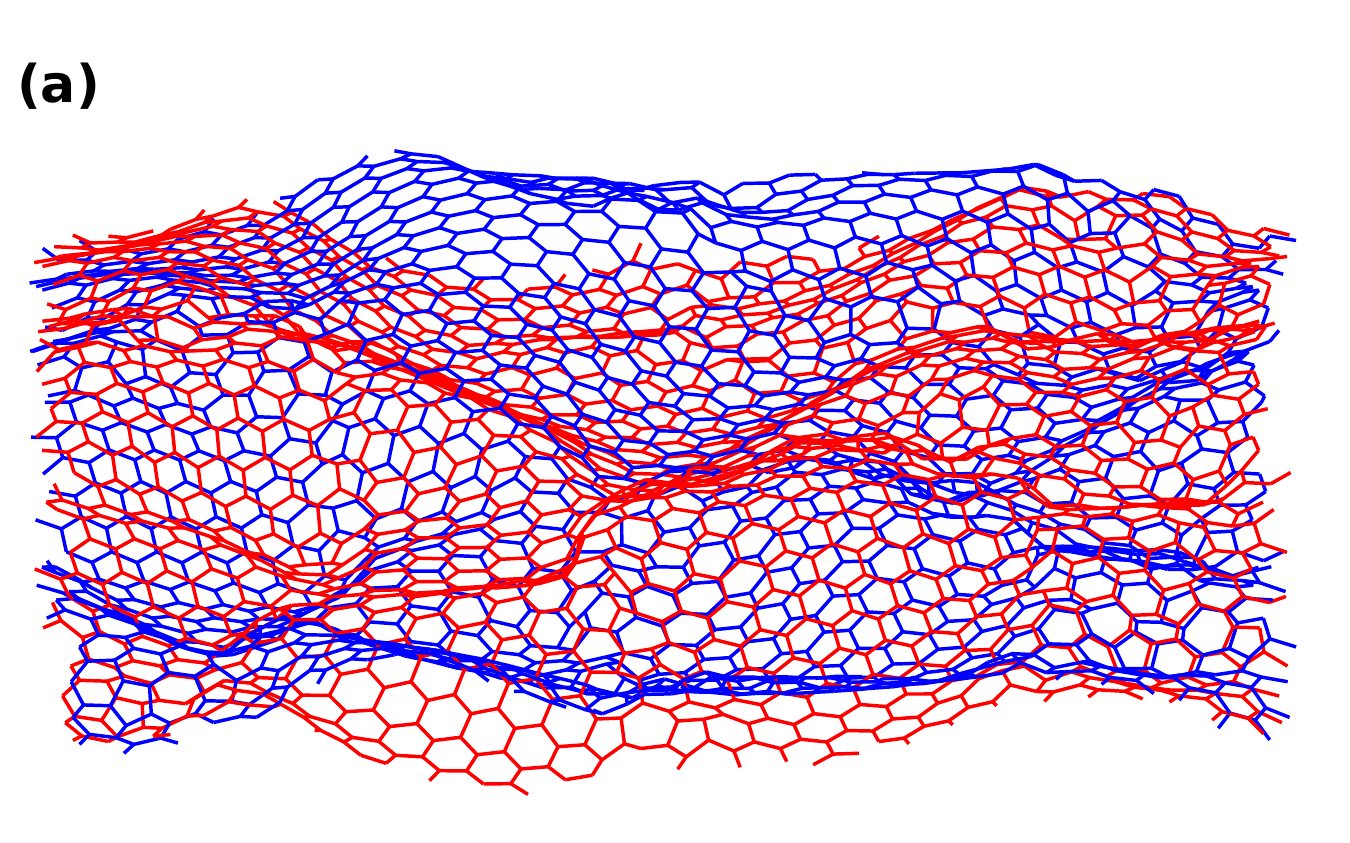}
    \end{subfigure}
    \begin{subfigure}[t]{0.45\textwidth}
        \centering
       \includegraphics[width=\textwidth]{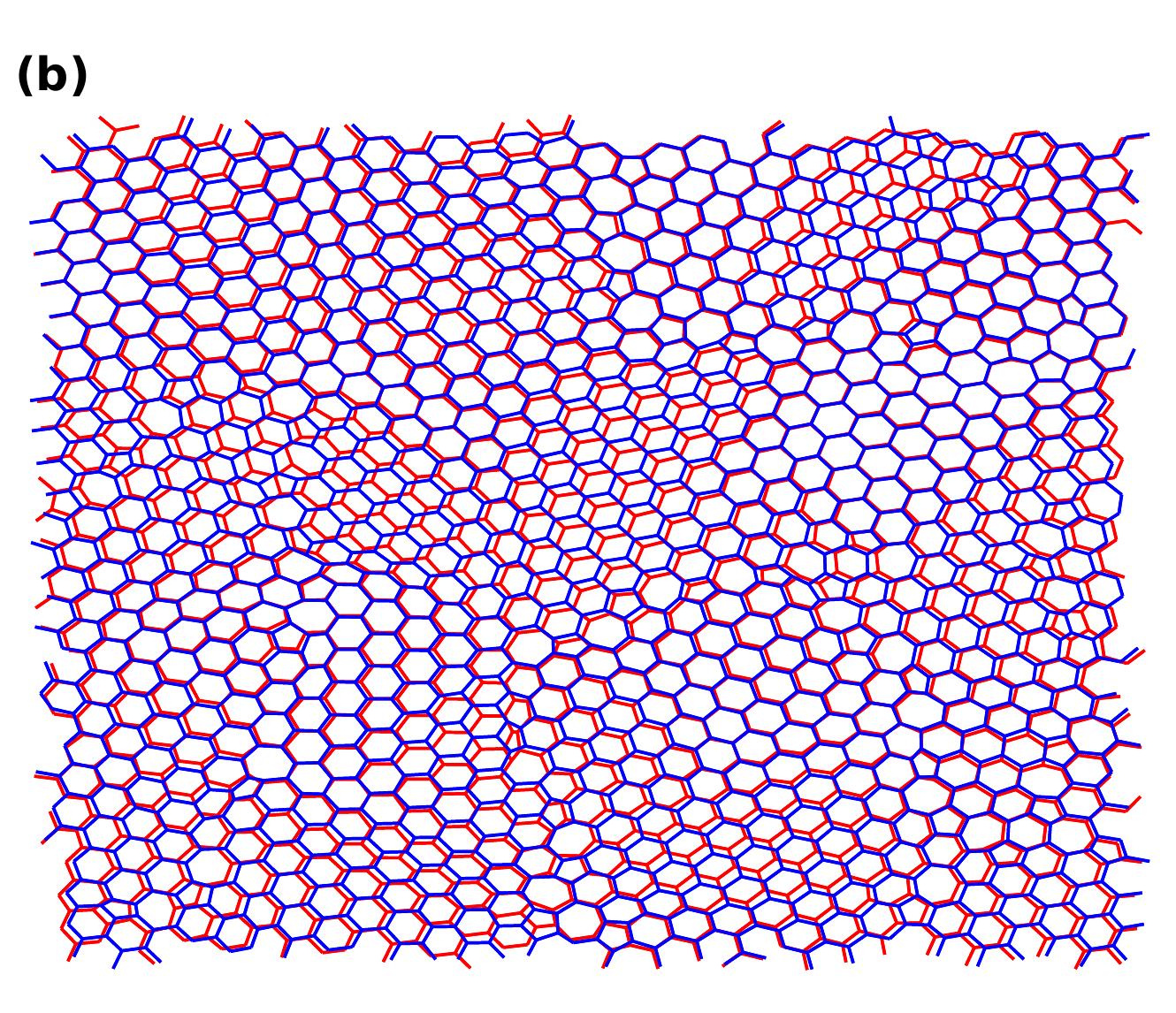}
    \end{subfigure}
    \caption{A sample of relaxed buckled graphene with $N=1600$ atoms, before and after a cycle in the configuration space at fixed topology. This is shown both from the side (a) and from the top (b). Its hysteretic behavior is clearly visible: the states at the beginning and at the end of a cycle are dramatically different. The energies of the two states are \SI{70.90}{\electronvolt} (red) and \SI{71.63}{\electronvolt} (blue). The depicted sample has been selected because the hysteretic behavior is particularly visible. }
    	\label{fig:two-samples}
\end{figure}

\emph{Numerical simulations and results.}
We first adiabatically stretch and relax a sample with $N=1600$ atoms in the way we previously described. After completing a cycle in the parameter space, the sample is in a different configuration, as it can be seen in Figure \ref{fig:two-samples}. By repeating this procedure a limited number of times, our sample ends up in a stable configuration. After repeating the cycle, the system reaches again a stable configuration, which is, however, different than the initial one.

The same procedure is then applied to a large sample containing $N=3200$ atoms but in this case we further characterize the sample during a gradual application of the external stress. 

\begin{figure}[t]
	\centering
	\includegraphics[height=7cm, width=9cm]{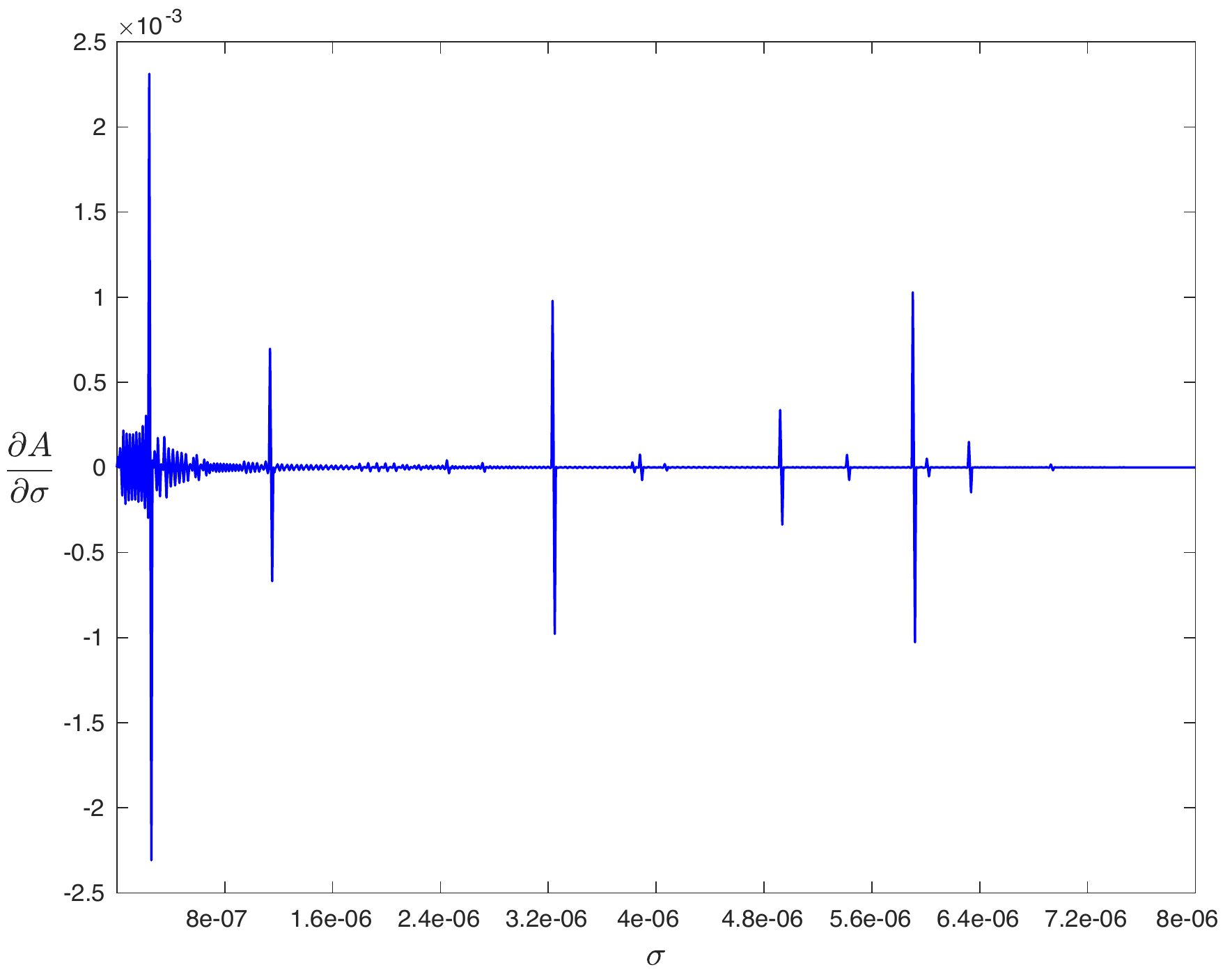}
	\caption{Evolution of the non-affinity parameter as the stretching force is gradually applied to a sample of relaxed graphene with $N=3200$ atoms at fixed topology. The evolution of the shape of the sample during the straining is discontinuous, as occasionally a tiny increase in stretching force causes a significant displacement, which manifests through the discontinuity of the derivative of the non-affinity parameter $A$, with respect to the applied stress $\sigma$, analogous to avalanches.}
	\label{fig:der}
\end{figure}

We start with a stable configuration and perform a cycle during which we compute the non-affinity parameter, defined in Eq.~(\ref{eq:non-affine-parameter}) and its derivative with respect to the stretching parameter. Figure \ref{fig:der} shows that the evolution of the sample during the straining is clearly discontinuous. Significant changes in the configuration are caused by small increases in stretching force, which results in  discontinuities in the derivative of the non-affine parameter.

\begin{figure}[t]
    \begin{subfigure}[t]{0.5\textwidth}
        \centering
        \includegraphics[height=6.8cm]{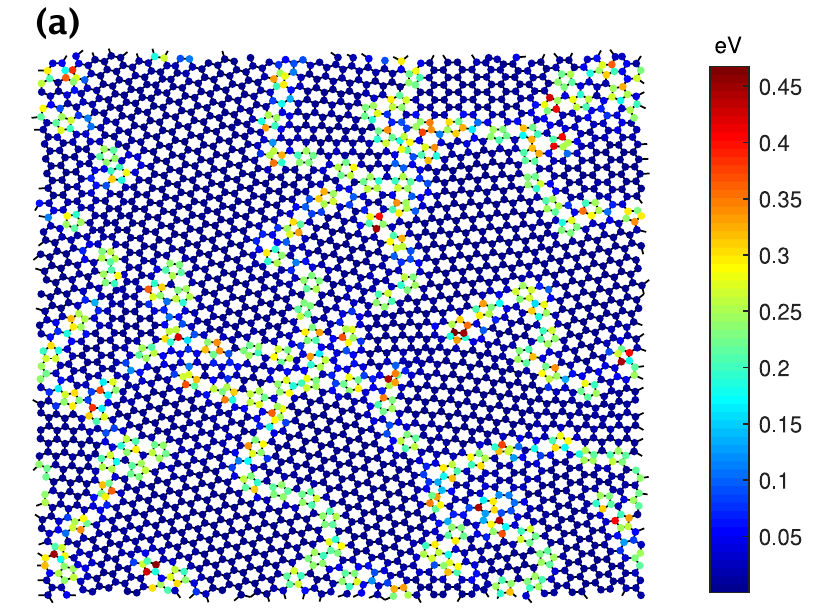}
    \end{subfigure}
    \begin{subfigure}[t]{0.5\textwidth}
        \centering
       \includegraphics[height=6.8cm]{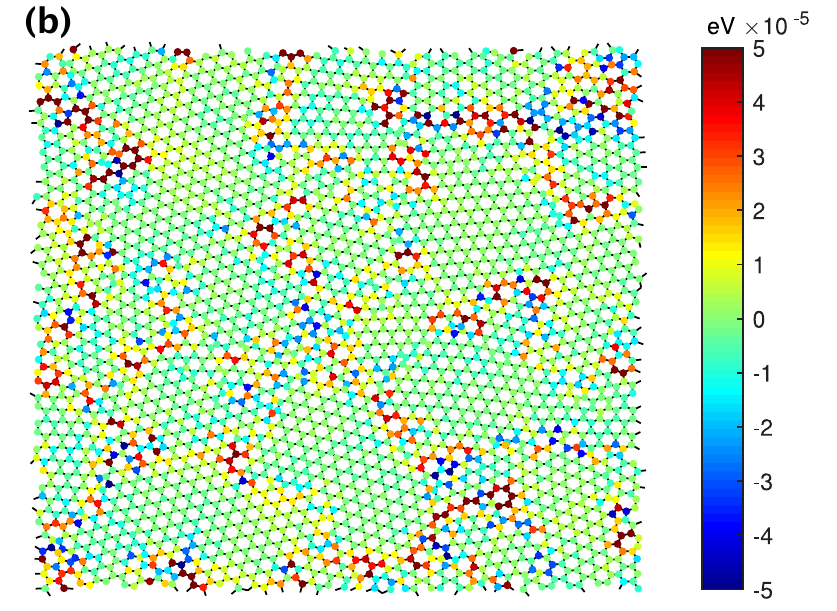}
    \end{subfigure}
    \caption{(a) Distribution of local energy in the graphene sample before a transition;  (b) the local energy difference before and after the transition during stretching. The graphene sample contains $N=3200$ atoms. The figure shows that the changes in the local energy distribution due to the transition are concentrated on and around ridges and vertices.}
    \label{fig:energy}
\end{figure}

We now analyze in more detail these discontinuities, in particular how the local energy distribution changes. As we can see in Fig.~\ref{fig:energy}, the change of energy is not uniform in the sample. The energy difference, both positive and negative, is concentrated on and around the ridge defects in the sample \cite{Witten2007,Fasolino2007}.
Separating the different energy contributions, we notice that, at each transition during the straining, the sample increases its internal energy $E_0$ given by Eq.~(\ref{eq:elastic-energy}), even though not all the terms contributing to it are necessarily positive. Nevertheless, the total elastic energy decreases during these transitions as the contribution of the stretching term, given by Eq.~ (\ref{eq:ESigma}) is negative, indicating a sudden increase of the surface area of the sample. This is confirmed by subtracting the affine component corresponding to the surface expansion or contraction during the transition, computed through a fit of the data points immediately preceding the transition, which show a significant non-affine component (see Supplemental Material at [URL will be inserted by publisher for a breakdown of the affine and non-affine components of the surface changes during the transition).  Finally, a minimal size of the sample in which this effect can be observable in our numerical experiment can be estimated as follows.  
In Fig.~ 2 the highest peak is $2.7\times10^{-3}$, while the threshold for the observation of the discontinuity is $0.4\times10^{-3}$. Given that the sample we used in our simulations (Fig.~3) contains about 15 grains, and assuming a linear scaling of the size of the discontinuity with the number of grains, yields an estimate of  a critical size of about 3 crystalline grains.

\emph{Discussion \& outlook.} In summary, we established that the evolution of graphene sheets is discontinuous under gradually applied external stress. It proceeds through a series of avalanche-like processes in which the ridges and vertices are created and annihilated, with the energy concentrated in these defects. Furthermore, the behavior of the graphene membrane is hysteretic: the system does not follow the same path back in the configuration space.
Our results imply that if twisting is applied gradually to graphene bilayers and multilayers, the change of the shape of such structures should be discontinuous in nature, which could be possibly relevant to recent experiments \cite{cao2018correlated,Cao2018,Yankowitz2019,Lu2019}. Our results should also be pertinent to other two-dimensional polycrystalline materials, including phosphorene, MoS$_2$ and hexagonal boron nitride (h-BN) whose experimental realization has been recently reported in Refs. \cite{Pei2016,Zhang2014,Xie2017}, as well as in van der Waals heterostructures \cite{Geim2013}. A continuous application of stress to a graphene elastic membrane necessarily involves a creation and annihilation of the topological defects in the form of ridges and vertices, which can be consequential for the electronic and mechanical properties of these systems. On the other hand, our findings should motivate experimental studies of the dynamical elasticity in these systems, in particular the evolution of the elastic moduli, vibrational density of states and thermal conductivity with the adiabatically applied external stress.
Finally, although we considered a specific case of monolayer graphene, we expect that our findings will be applicable to generic membranes embedded in three-dimensional space. In particular, within this approach it would be interesting to study the dynamical properties of origami metamaterials~\cite{Santangelo2017}.

\bibliographystyle{apsrev4-1}
\bibliography{references}
\end{document}